\documentclass{elsart3}
\usepackage{graphicx}
\usepackage{amssymb}
\begin{document}
\begin{frontmatter}

\title{New frequency-modulation readout based on relaxation oscillations}
\author{Miha Furlan\corauthref{cor}},
\ead{miha.furlan@psi.ch}
\author{Eugenie Kirk, and Alex Zehnder}
\corauth[cor]{Corresponding author. Tel.: +41-56-310-4519.}

\address{Paul Scherrer Institute, Laboratory for Astrophysics,
5232 Villigen PSI, Switzerland}

\begin{abstract}
Scaling of multi-pixel cryogenic detectors for imaging
becomes increasingly difficult with size due to complexity of
readout circuitry and cryogenic constraints (thermal load from wiring).
We propose and demonstrate a new readout scheme based on a
highly stable RF oscillator composed of a superconducting
tunnel junction which exhibits relaxation oscillations.
The oscillation frequency is almost linear with the analog bias signal
over a wide operation range. The frequency signals
from different detectors can be combined into one single readout line.
The current noise of an optimized circuit is about
$5\, \mathrm{pA}/\sqrt{\mathrm{Hz}}$, which is
comparable to standard SQUID amplifiers.
We show experimental data from `stand-alone' operation as well as
response to microcalorimeter X-ray signals.
\end{abstract}

\end{frontmatter}

Cryogenic radiation detectors
have proven to be the devices of choice when
highest energy sensitivity
(in the X-ray range),
single photon detection efficiency
and direct spectroscopic resolution
are required.
For astrophysical observations, a trend toward
large arrays for imaging has been
accompanied by efforts to solve the non-trivial
problem of cryogenic multiple-channel readout.
While most readout schemes are based on
multiplexed SQUID amplifiers,
we propose and demonstrate an alternative
and relatively simple low-noise
analog-to-frequency conversion circuit.

\begin{figure}[b]
   \centering
   \includegraphics[width=0.82\linewidth]{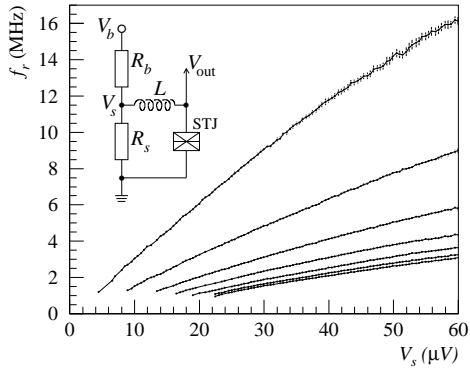}
   \caption{
   Measured $f_r$ as a function of $V_s$ and for different
   $I_c$ (modified by application of magnetic field).
   Fitting the theoretical $f_r$ to the data points yielded
   (from bottom to top):
   $I_c = 58.07, 53.71, 48.15, 39.55, 29.98, 19.69,$ and
   $10.63 \, \mu \mathrm{A}$ ($I_c^0 = 58.3 \, \mu \mathrm{A}$).
   The circuit parameters were: $L = 280\, \mathrm{nH}$,
   $R_s = 91 \, \mathrm{m}\Omega$, and the STJ having
   $V_g = 330 \mu \mathrm{V}$ and $R_n = 1.2 \, \Omega$. }
   \label{FvsV.fig}
\end{figure}

The operating principle is based on a hysteretic superconducting
tunnel junction (STJ) exhibiting relaxation
oscillations \cite{Vernon1968} in the RF range.
If the gaps $\Delta_1$ and $\Delta_2$ of the two
superconducting electrodes differ slightly
($0 < \vert \Delta_1 - \Delta_2 \vert \ll \Delta_1 , \Delta_2$),
the current-voltage characteristics of the device show a
region with negative differential resistance where
biasing of the device is potentially unstable \cite{Albegova1969}.
An oscillator circuit can be built with an STJ (normal
resistance $R_n$, capacitance $C_j$ and critical current $I_c$)
in series with an inductor $L$, and voltage biasing both at
$V_s$ by a shunt $R_s \ll R_n$, as depicted in the inset
of Fig.~\ref{FvsV.fig}.

Operating the circuit at a current $I_b > I_c$ determined by the
current limiting resistor $R_b \gg R_s$ the relaxation oscillations
are dominated by \cite{Vernon1968,FurlanRelax}
\begin{itemize}
\item the current rise time
\begin{equation} \label{tau_sc.eq}
\tau_{sc} = - \, \frac{L}{R_s} \ln \left( 1 - \alpha^{-1} \right)
\end{equation}
on the supercurrent (SC, zero voltage) branch of the
current-voltage curve, where $\alpha = I_b / I_c$ is the
normalized bias parameter, and
\item the current decay time
\begin{equation} \label{tau_qp.eq}
\tau_{qp} = \frac{L}{R_s + R_{qp}} \ln
\left( 1 + \frac{(R_s + R_{qp}) I_c}{V_g - V_s} \right)
\end{equation}
on the quasiparticle (QP) branch at the gap voltage
$V_g = (\Delta_1 + \Delta_2 )/e $, where
$R_{qp}$ is the junction resistance in the gap region.
\end{itemize}
The voltage switching times between SC and
QP branch are on the order of
$\tau_v  \approx C_j V_g / I_c$
and are usually negligibly fast.
For $V_s \ll V_g$ and $R_{qp} \gg R_s$ we find
$\tau_{sc} \gg \tau_{qp}$ which yields an approximation
for the relaxation oscillation frequency:
\begin{eqnarray} \label{tau_r.eq}
f_r & = & \frac{R_s}{L} \left( \alpha - \frac{1}{2} -
\mathrm{O} \left( \alpha^{-1} \right) \right) \nonumber \\
    & = & \frac{V_s}{I_c L} \, ,  \qquad (\alpha \gg 1) \,  .
\end{eqnarray}
Hence, $f_r$ is proportional to $V_s$
(or to $I_b = V_s / R_s$) and the circuit behaves
as an almost linear analog-to-frequency converter
over an acceptably broad dynamic range.

Relaxation oscillations have been measured at
$100 \, \mathrm{mK}$ for various circuit parameters.
Experimental $f_r (V_s)$ dependence for
different critical currents $I_c = I_c^0 / \kappa$ is shown in
Fig.~\ref{FvsV.fig}, where the nominal value $I_c^0$
may be suppressed by a factor $\kappa$ to the effective
$I_c$ due to application of a magnetic field. This
is particularly convenient for tuning of the
circuit behaviour or eventually extending the
operation range to lower currents.
The deviation from linearity observed in Fig.~\ref{FvsV.fig}
is fully accounted for by using both
Eqs.~(\ref{tau_sc.eq}) and (\ref{tau_qp.eq}),
because $\tau_{qp}$ starts to contribute at higher $V_s$.

\begin{figure}[h]
   \centering
   \includegraphics[width=0.80\linewidth]{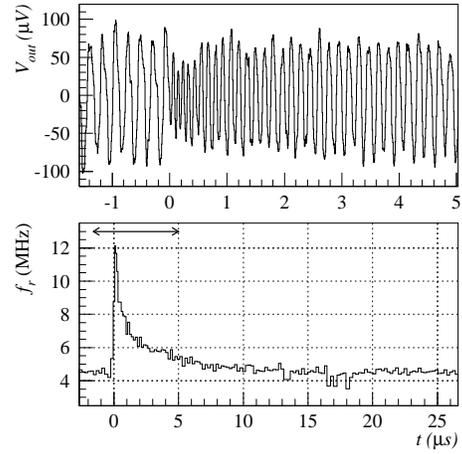}
   \caption{
   Top: Relaxation oscillations during a SINIS
   detector $6 \, \mathrm{keV}$ X-ray event
   (the detector replacing $R_b$).
   The amplitude modulation and sinusoidal oscillation are
   due to the externally applied band-pass filter.
   Bottom: Time sequence of inverse oscillation periods,
   equivalent to a time-dependent $f_r$,
   extracted from the above analog signal
   (note the larger time scale, while the arrow indicates the
   range of the top graph).
   Circuit and device parameters were: $L = 48\, \mathrm{nH}$,
   $R_s = 91 \, \mathrm{m}\Omega$, $I_c = 7.28 \, \mu \mathrm{A}
   \, (\kappa = 8)$. }
   \label{Xray.fig}
\end{figure}

If we now introduce a cryogenic detector by replacing $R_b$
with a high-resistance device (e.g. a NIS microcalorimeter) or
$R_s$ with a low-resistance device (e.g. a TES), the
detector signal will be directly converted to a change in $f_r$.
The oscillator signal $V_\mathrm{out}$ with relatively large
amplitude $V_g$ is demodulated with conventional
(phase-locked loop) electronics outside the cryostat.
The measured response of the oscillator to a SINIS detector
\cite{FurlanSINIS}  X-ray event is shown
in Fig.~\ref{Xray.fig}.

The detector was biased close to its gap voltage
($V_d = 2 \Delta_\mathrm{Al} /e$)  at
$V_b = 334 \, \mu \mathrm{V} = 0.98 \, V_d$
for operation at sufficiently high dark current
$I_b^0 = 17.5  \, \mu \mathrm{A} = 2.4 \, I_c$.
The peak current of the measured analog signal
was $I_b^1 = 46.4 \, \mu \mathrm{A}$ (the indices $0$ and $1$
denote the dark and irradiated detector states, respectively).
The corresponding frequencies extracted from Fig.~\ref{Xray.fig}
are $f_r^0 = 4.47 \, \mathrm{MHz}$ and
$f_r^1 = 12.1 \, \mathrm{MHz}$, which, using
Eq.~(\ref{tau_r.eq}), give currents of
$17.2  \, \mu \mathrm{A}$ and $46.5  \, \mu \mathrm{A}$,
respectively. There is excellent agreement between analog
and frequency-modulated signals, which is particularly remarkable
considering the model simplifications and experimental uncertainties.
The relative width
$\langle \delta \tau_r \rangle / \langle \tau_r \rangle$
of the oscillation period distribution (in `stand-alone'
operation without noisy detector) was typically on the order of
$10^{-2}$.

For observation of stable relaxation oscillations and optimum noise
behaviour, several constraints on circuit parameters have to be
considered \cite{FurlanRelax}, a discussion of which is beyond
the scope of this paper.
Nevertheless, we present calculations and realistic estimates
for the proposed circuit including standard
(SI)NIS and TES detectors.

The main noise sources from
the oscillator are Johnson noise from $R_s$, shot noise from
tunneling, $1/f$ flicker noise from two-level fluctuators
\cite{Wellstood2004} and thermally activated escape from the
zero-voltage state \cite{Fulton1974}.
For typical STJ devices and high-frequency operation
the process of
thermal excitation over an energy barrier was found
to dominate the circuit noise. Depending on junction parameters,
temperature $T$, and current slew rate
$\d I / \d t \approx I_c / \tau_{sc}$
the transitions from the superconducting to the resistive states
will occur at a current $I_m < I_c$. The width
$\langle \delta I_m \rangle / I_c$
of the transition current probability
function is, to first order, proportional to $(T / I_c)^{2/3}$
\cite{Snigirev1983}.
Critical-current fluctuations are a direct linear cause for
fluctuations of the relaxation oscillation period
$\delta \tau_r / \tau_r = \delta I_m / I_m$, and we can
determine a current noise density (referred to the
circuit input and valid over a limited parameter range
\cite{FurlanRelax}) of
\begin{equation}
j_b  =  \frac{I_b}{\sqrt{f_r}} \frac{\delta \tau_r}{\tau_r}
     = \frac{\alpha I_c}{\sqrt{f_r}} \frac{\delta I_m}{I_m}
     \propto \frac{\alpha I_c^{1/3} T^{2/3}}{f_r^{1/2}} \, .
\end{equation}

Table~\ref{params.tab} summarizes calculated optimum parameters for
realistic (SI)NIS and TES readout.
For the current-to-voltage converter (NIS readout) we have chosen
the smallest STJ dimensions which can still be easily fabricated
by standard optical lithography. In the case of TES readout we are
restricted to a typical resistance $R_s$ of the detector.
The current-noise levels $j_b$ in Table~\ref{params.tab} are
sufficiently low compared to detector or SQUID noise.

The most apparent advantages of the relaxation oscillation based
readout compared to existing technologies are that it neither
requires narrow band filtering as in
frequency-domain multiplexing, nor does it suffer from
switching complications as in time-domain multiplexing.
Furthermore, there is no need for complex bias and control circuitry.
The main disadvantage of our TES readout scheme is the operation
in the less favoured and potentially unstable current biasing mode.
Finally, this scheme
is probably not appropriate for readout of STJ
detectors due to their intrinsically low current levels
compared to $I_c$ of the oscillator STJ.

\begin{table}
\caption{Examples of optimum STJ and circuit parameters for
readout of NIS and TES detectors, where $\ell$ is the STJ side
length (see main text for the other parameters).
In the TES case $R_s$ corresponds to the detector itself
with a typical operating point resistance.
Both circuits are operated at $T = 100 \, \mathrm{mK}$
and $\alpha = 3$.}
\label{params.tab}
\begin{tabular}{l c c c c c c c c c}
\hline
 & $\ell$ & $R_n$ & $\kappa$ & $I_c$ &
   $L$ & $R_s$ & $f_r$ & $\frac{\delta \tau_r }{\tau_r}$ & $j_b$ \\
 & $(\mu \mathrm{m})$ & $(\Omega)$ & $\qquad$ &
   $(\mu \mathrm{A})$ & $(\mu \mathrm{H})$ &
   $\, (\Omega) \,$ & $(\mathrm{MHz})$ & $\, ( \cdot 10^3 ) \,$ &
   $\left( \frac{\mathrm{pA}}{\sqrt{\mathrm{Hz}}} \right)$ \\
\hline
NIS & 5  & 40  & 12 & 0.52 & 0.8 & 8   & 30  & 17  & 5.1 \\
TES & 20 & 2.5 & 15 & 6.68 & 0.1 & 0.1 & 3   & 3.1 & 37  \\
\hline \\
\end{tabular}
\end{table}

\section*{Acknowledgements}
We are indebted to Ph.\ Lerch for valuable discussions
and to F.\ Burri for technical support.

\end{document}